\documentclass[12pt]{article}
\usepackage{epsf,epsfig,amsmath}
\begin{document}

\title{The Bell Theorem as a Special Case of a Theorem of Bass}

\author{Karl Hess$^1$ and Walter Philipp$^2$}

\date{$^1$ Beckman Institute, Department of Electrical Engineering
and Department of Physics,University of Illinois, Urbana, Il 61801
\\ $^{2}$ Beckman Institute, Department of Statistics and Department of
Mathematics, University of Illinois,Urbana, Il 61801 \\ }
\maketitle

\begin{abstract}

The theorem of Bell states that certain results of quantum
mechanics violate inequalities that are valid for objective local
random variables. We show that the inequalities of Bell are
special cases of theorems found ten years earlier by Bass and
stated in full generality by Vorob'ev. This fact implies precise
necessary and sufficient mathematical conditions for the validity
of the Bell inequalities. We show that these precise conditions
differ significantly from the definition of objective local
variable spaces and as an application that the Bell inequalities
may be violated even for objective local random variables.

\end{abstract}

\section{Introduction}

Einstein-Podolsky-Rosen (EPR) \cite{einstein} suggested that
quantum mechanics was incomplete and that hidden variables may be
needed for completion. This was the subject of an extensive debate
with Bohr \cite{bohr} who denied the existence of such hidden
variables using his well known reasoning that is at the basis of
the Copenhagen interpretation of quantum mechanics. A mathematical
non-existence proof for these hidden variables was presented by
von Neumann within the framework of quantum mechanics. Bell
\cite{bellbook}, however, showed that von Neumann's proof assumed
the simultaneous measurability of certain quantities that could
not possibly be simultaneously measured. Subsequently Bell himself
presented a non-existence proof \cite{bell} in form of
inequalities that are derived by using more complex assumptions
that did not necessarily include simultaneous measurability.

The inequalities of Bell \cite{bell} are derived by the use of
probability theory, in essence by the use of very elementary facts
about random variables within the framework pioneered by
Kolmogorov. The derivation of the inequalities does not involve
physics or quantum mechanics, yet the inequalities have assumed an
important role for the foundations of quantum mechanics. This role
is a consequence of the assumptions for the random variables (and
other possible variables) that are used to derive the Bell
inequalities. It is commonly believed that these assumptions are
needed and moreover are equivalent to the basic postulates of an
``objective local" parameter space \cite{leggett}. These
postulates or conditions encompass Einstein-locality and the
existence of elements of reality in the sense of Mach that
co-determine the outcome of measurements. The fact that some
results of quantum mechanics violate the Bell inequalities has
therefore led to the belief that no objective local parameter
space can exist that explains all the results of quantum
mechanics.

Subsequent to these discussions experiments were proposed
\cite{ba}, \cite{bohm}  and later realized \cite{eprex} that
confirmed the theoretical result of quantum mechanics. This seemed
to leave only difficult options for theoretical physics such as
(i) to deny that the microscopic entities of physics have
objective reality (non-existence of objective local parameters) or
(ii) to assert that an influence can be propagated faster than the
speed of light \cite{moore}. Additional options involving the
validity of counterfactual reasoning have also been suggested and
will be discussed below.

We show in this paper that because of the historical sequence of
events, viz. the development of Bell's theory before the
performance of the actual experiments, some very important facts
of probability theory have not been considered and/or
misinterpreted . These facts are connected with the concept of a
probability space that is important to link probability theory and
mathematical statistics to the evaluation of actual experiments.
We reconsider here these concepts and show that violations of the
Bell inequalities have a purely mathematical reason, in particular
that the Bell inequalities represent a special case of theorems
given earlier by Bass \cite{bass}, Vorob'ev \cite{vorob1},
\cite{vorob} and Schell \cite{schell}. These theorems permit us to
deduce that, for all the possible Bell inequalities to be valid,
it is a necessary and sufficient condition that the random
variables involved in their proof are defined on one common
probability space. Beyond this we show that the requirement of the
use of one common probability space does not follow from the
requirements of objective local spaces and vice versa. In fact, we
show that there exist objective local random variables that can
not be defined on a common probability space and therefore do not
need to obey the Bell inequalities. As mentioned, Bell has
dismissed von Neumann's proof that assumed from the start the
simultaneous measurability of the observables that correspond to
our random variables and that can not be measured simultaneously.
Our contribution here is that we also dismiss non-existence proofs
of certain systems of random variables that need to be defined on
one common probability space when it is clear from the outset that
these systems of random variables can not be defined on any common
probability space. We believe that our results give additional
options of explanation for Aspect-type experiments without
violating relativity or denying objective reality.

\section{Mathematical model of a singlet spin state EPR
experiment}

As outlined above, probability theory provides a probability space
and random variables and the link to the statistical treatment of
the data from an actual experiment. Bell \cite{bell} considered an
experimental situation advocated by Bohm and Aharonov \cite{ba},
and by Bohm and Hiley \cite {bohm}. This proposal was transformed
into an actual experiment by Aspect et al. \cite{eprex} including
the suggestion of Bell and others that a rapid change of the
settings needed to be implemented to accomplish a delayed choice
situation \cite{eprex}. We develop now an idealized experiment and
a probability model for this actual experiment by Aspect et al
within the framework of Kolmogorov. We note that our procedure
also applies to other related experiments.

We first recall the concepts of a (discrete) probability space and
of a random variable defined on it.  As Feller states
\cite{feller}``If we want to speak about experiments or
observations in a theoretical way and without ambiguity, we first
must agree on the simple events representing the thinkable
outcomes; {\it they define the idealized experiment}...... By
definition {\it every indecomposable result of the (idealized)
experiment is represented by one, and only one, sample point}. The
aggregate of all sample points will be called the {\it sample
space}." In our case of the idealized EPR experiment, the simple
event can be chosen, for example, as the event of sending out one
(and only one) correlated pair. With this event we associate an
element $\omega$. In order to avoid mathematical technicalities
that are not needed for the purpose of our paper we assume that
the sample space $\Omega$ is at most countable. Each simple event
$\omega \in \Omega$ is assigned the probability $P(\omega)$ that
$\omega$ occurs. $P$ is a set function, defined for all subsets of
$\Omega$, that satisfies the usual axioms, such as countable
additivity and it assigns to $\Omega$ the value $P(\Omega) = 1$.
The pair $(\Omega, P)$ is called a probability space. A random
variable is a real-valued function on $\Omega$, but if needed it
also can assume values in high-dimensional space.

We now turn to the specifics of an idealized EPR-experiment. A
correlated spin pair in the singlet state is sent out from a
source in opposite directions toward measurement stations. These
stations are characterized by certain randomly and rapidly
switched settings which we denote by three dimensional unit
vectors ${\bf a}, {\bf b}, {\bf c}, ...$. The measurements in the
stations are mathematically represented by random variables $A =
\pm 1, B = \pm 1, C = \pm 1, ...$ that may in turn be functions of
other random variables e.g. a source parameter $\Lambda$ that
characterizes all the properties of the particles sent out from
the source. $A$ indicates that the measurements that correspond to
the outcomes of random variable $A$ have been performed using the
setting $\bf a$ and similarly for $B$ and $C$.

We perform three categories of experiments, each with a different
pair of setting vectors. The first category is characterized by
the vectors $\bf a$ in station $S_1$ and $\bf b$ in station $S_2$.
According to our notational convention we denote the pair of
measurements $(A, B)$ and the joint probability density of $A$ and
$B$ by $f_1$. Thus $f_1$ is given by
\begin{equation}
f_1 (+1, +1) = P(A=+1, B=+1) \text{   ,   }f_1(-1, +1) = P(A=-1,
B=+1) \nonumber
\end{equation}
\begin{equation}
f_1 (+1, -1) = P(A=+1, B=+-1) \text{   ,   }f_1(-1, -1) = P(A=-1,
B=-1) \label{bvc1}
\end{equation}
The second category of experiments will be characterized by the
vectors $\bf a$ in $S_1$ and $\bf c$ in $S_2$ with the resulting
pair of measurements $(A, C)$ having density $f_2$, and the third
category by the vectors $\bf b$ in $S_1$ and $\bf c$ in $S_2$
resulting in a pair of measurements $(B, C)$ with density $f_3$.

The measurement outcomes on both sides need to be completely
random and $\pm 1$ with equal probability, i.e. all three
distributions have identical marginals. This is dictated by the
rules of quantum mechanics and verified by experiment. From this
it follows that the $f_i, i =1, 2, 3$ have the center of gravity
for their point masses at the origin $(0, 0)$ and
\begin{equation}
f_i(+1,-1) + f_i(+1,+1) = \frac {1} {2} = f_i(+1,+1) +
f_i(-1,+1)\text{   for   } i = 1,2,3 \label{bv1}
\end{equation}

The idealized mathematical model with exactly the properties
described above and used within the framework of Kolmogorov is the
basis for all our further considerations and we call it the Ma-EPR
model.

\section{Ma-EPR and the theorems of Bass and Vorob'ev}

We start with an example that illustrates the theorems of Bass
\cite{bass} and Vorob'ev \cite{vorob} for the special case of the
Ma-EPR model. The essential point of these theorems is that, in
general, it is not possible to find three random variables $A, B$
and $C$, defined on a common probability space such that the three
pairs $(A,B)$, $(A,C)$, and $(B,C)$ of random variables have their
joint densities equal to $f_1, f_2$ and $f_3$, respectively. Hence
the notation that is commonly used and that we also introduced in
section 2 above is misleading in the sense that it suggests there
exist three random variables $A, B$ and $C$ that can reproduce the
joint densities $f_1$, $f_2$ and $f_3$, when in fact they can not.
Here is a modification of an example of Vorob'ev \cite{vorob}.

\begin{table}[ht]
\centering
    \begin{tabular}{|l||r|r|r|r|}\hline
   & $(+1,+1)$ & $(+1,-1)$ & $(-1,+1)$ & $(-1,-1)$
\\ \hline
      $f_1(.,.)$ & $3/8$ & $1/8$
       & $1/8$ & $3/8$\\ \hline
      $f_2(.,.)$ & $3/8$ & $1/8$
       & $1/8$ & $3/8$\\ \hline
      $f_3(.,.)$ & $1/8$ & $3/8$
      & $3/8$ & $1/8$\\ \hline
   \end{tabular}
   \caption{Vorob'ev-type example \cite{vorob}.}\label{TA:ma}
\end{table}

Clearly Eq(\ref{bv1}) holds. Suppose now that three such random
variables $A, B$ and $C$ exist and are defined on one common
probability space. Then the first two rows would imply that $P(A =
B) = \frac {3} {4} = P(A = C)$, and so $P(B \neq C) \leq \frac {1}
{2}$ , contradicting the fact that according to the third row $P(B
= C) = \frac {1} {4}$. Another easy way to see that three such
random variables cannot be defined on a common probability space
follows from the fact that, for instance, it is not possible to
assign a probability to the event $(A = 1, B = 1, C = 1)$.
According to the first entry of the third row this probability
could not exceed $\frac {1} {8}$. Subtracting this value from the
first entry of the first row we obtain that P(A = 1, B = 1, C =
-1) would have to be at least $\frac {1} {4}$. But this is in
conflict with the second entry of the second row. The reason for
this phenomenon is that, picturesquely speaking, the three pair
distributions form a closed loop. The joint densities of $(A, B)$
and of $(A, C)$ already contain some information about the joint
density of $(B, C)$. Hence we do not have complete freedom to
choose the latter one. This was shown for three general pair
distributions by Jean Bass \cite{bass} and independently by Schell
\cite{schell} who also investigated the connection with certain
problems in economics. Vorob'ev \cite{vorob1}, \cite{vorob}
established necessary and sufficient conditions that any complex
of distributions must possess so that these distributions can be
realized as marginal distributions of a set of random variables
defined on a common probability space.

\begin{table}[ht]
\centering
    \begin{tabular}{|l||r|r|r|r|}\hline
   & $(+1,+1)$ & $(+1,-1)$ & $(-1,+1)$ & $(-1,-1)$
\\ \hline
      $f_1(.,.)$ & ${\frac {1} {4}}(1 + \sigma_1)$ & ${\frac {1} {4}}(1 - \sigma_1)$
       & ${\frac {1} {4}}(1 - \sigma_1)$ & ${\frac {1} {4}}(1 + \sigma_1)$\\ \hline
      $f_2(.,.)$ & ${\frac {1} {4}}(1 + \sigma_2)$ & ${\frac {1} {4}}(1 - \sigma_2)$
       & ${\frac {1} {4}}(1 - \sigma_2)$ & ${\frac {1} {4}}(1 + \sigma_2)$\\ \hline
      $f_3(.,.)$ & ${\frac {1} {4}}(1 + \sigma_3)$ & ${\frac {1} {4}}(1 - \sigma_3)$
      & ${\frac {1} {4}}(1 - \sigma_3)$ & ${\frac {1} {4}}(1 + \sigma_3)$\\ \hline
   \end{tabular}
   \caption{Pair densities in terms of covariances}\label{TA:ob}
\end{table}

It is easy to show that under the assumption of Eq(\ref{bv1}) the
joint pair densities can be expressed in terms of the covariances
$\sigma_i, i=1,2,3$ defined by these pair densities. The pair
densities are then given by Table \ref{TA:ob} (see also the Lemma
below). Note that the covariances $\sigma_i$ do not exceed $1$ in
absolute value. Suppose now that there exist three random
variables $A, B, C$ defined on one common probability space that
reproduce the densities $f_1, f_2, f_3$ in Table \ref{TA:ob}. Then
$\sigma_1 = E(AB)$, $\sigma_2 = E(AC)$, and $\sigma_3 = E(BC)$
where $E$ denotes the expectation value. Expressing the entries of
Table \ref{TA:ob} in terms of the eight unknown probabilities $P(A
= \pm 1, B = \pm 1, C = \pm1)$ will result in a system of twelve
linear equations in these eight unknowns that can be solved in an
elementary way. In particular, solving this system shows that
these eight probabilities can be expressed in terms of the three
covariances $\sigma_i, i = 1,2,3$. It turns out that five of these
twelve linear equations are redundant. Thus this system has
infinitely many solutions. Taking into account that the solutions
of this system represent probabilities $P \geq 0$ we obtain in a
straightforward way that the following four inequalities are
necessary and sufficient conditions for the solvability of the
consistency problem for the three pair distributions given in
Table \ref{TA:ob}:
\begin{equation}
1 + \sigma_1 + \sigma_2 + \sigma_3 \geq 0 \label{bvcc1}
\end{equation}
\begin{equation}
1 + \sigma_1 - \sigma_2 - \sigma_3 \geq 0 \label{bvcc2}
\end{equation}
\begin{equation}
1 - \sigma_1 + \sigma_2 - \sigma_3 \geq 0 \label{bvcc3}
\end{equation}
\begin{equation}
1 - \sigma_1 - \sigma_2 + \sigma_3 \geq 0 \label{bvcc4}
\end{equation}

Of course, the necessity part of this conclusion can be shown
directly and trivially by modifying the standard proofs of the
Bell inequality along the lines shown in \cite{bell}.

\section{Bell's inequalities as a special case of Bass-Vorob'ev}

Replacing the covariances $\sigma$ by the corresponding
expectation values, one obtains from
Eqs.(\ref{bvcc2}-\ref{bvcc3}):
\begin{equation}
E(AB) - E(AC) \leq 1 - E(BC) \label{bv6}
\end{equation}
and
\begin{equation}
-E(AB) + E(AC) \leq 1 - E(BC) \label{bv7}
\end{equation}
Eqs.(\ref{bv6}) and (\ref{bv7}) give
\begin{equation}
|E(AB) - E(AC)| \leq 1 - E(BC) \label{bv8}
\end{equation}

This is, of course, one of the celebrated Bell inequalities. Five
more can be obtained in analogous fashion from
Eqs.(\ref{bvcc1}-\ref{bvcc4}) giving a total of 6 (4 choose 2).
These can also be obtained by cyclic permutation in Eq.(\ref{bv8})
and replacing both minus signs by plus signs.

\begin{figure}[htbp]
    \centering
        \includegraphics[width=0.30\textwidth]{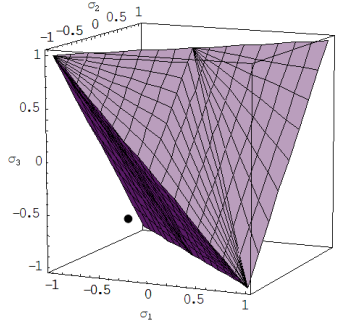}
    \caption{Covariance Tetrahedron. The solid point represents a
     choice of values that violates the Bell inequalities.}
    \label{fig:Tetrahedra}
\end{figure}

Bass \cite{bass} proved that for three general pair distributions
the consistency problem can be solved if and only if the triple
$(\sigma_1, \sigma_2, \sigma_3)$ considered as a point in $R^3$
belongs to a certain domain. In the special case we have been
considering this domain reduces to the tetrahedron defined by the
inequalities of Eqs.(\ref{bvcc1}-\ref{bvcc4}). We shall call it
the covariance tetrahedron. It is diplayed in Fig.
\ref{fig:Tetrahedra}.

We formulate now our findings for the Ma-EPR experiment as a
theorem. We first collect a few facts of section 3 above in form
of a lemma.

Lemma: Let $f$ be a density supported on the four vertices $(\pm
1, \pm 1)$ of a square. Suppose that
\begin{equation}
f(+1, +1) + f(+1, -1) = f(+1, +1) + f(-1, +1) = \frac {1} {2}
\label{bvcc5}
\end{equation}
Then
\begin{equation}
\sum x f(x, y) = \sum y f(x, y) = 0 \label{bvcc6}
\end{equation}
where the sums are extended over the four points $(x, y) = (\pm 1,
\pm 1)$. Conversely, if $f$ satisfies Eq.(\ref{bvcc6}) then $f$
also satisfies Eq.(\ref{bvcc5}).

Set
\begin{equation}
\sigma := \sum x y f(x, y) \label{bvcc7}
\end{equation}
with the same proviso for the sum. Then $f$ can be expressed in
terms of $\sigma$ by the equations
\begin{equation}
f(+1, +1) = f(-1, -1) = {\frac {1} {4}}(1 + \sigma) \label{bvcc8}
\end{equation}
\begin{equation}
f(-1, +1) = f(+1, -1) = {\frac {1} {4}}(1 - \sigma) \label{bvcc9}
\end{equation}

Theorem1: Let $f_1, f_2, f_3$ be three probability densities
satisfying the hypotheses of the Lemma with corresponding
covariances $\sigma_1, \sigma_2, \sigma_3$. Then the following
statements are equivalent

\begin{enumerate}

\item[(I)] The point $(\sigma_1, \sigma_2, \sigma_3) \in R^3$
satisfies the system of inequalities Eqs.(\ref{bvcc1}-\ref{bvcc4})
and therefore belongs to the covariance tetrahedron.

\item[(II)] The point $(\sigma_1, \sigma_2, \sigma_3)$ satisfies
the following six Bell-type inequalities
\begin{equation}
|\sigma_1 - \sigma_2| \leq 1 - \sigma_3 \text{   ,   }|\sigma_1 +
\sigma_2| \leq 1 + \sigma_3 \nonumber
\end{equation}
\begin{equation}
|\sigma_1 - \sigma_3| \leq 1 - \sigma_2 \text{   ,   }|\sigma_1 +
\sigma_3| \leq 1 + \sigma_2 \nonumber
\end{equation}
\begin{equation}
|\sigma_2 - \sigma_3| \leq 1 - \sigma_1 \text{   ,   }|\sigma_2 +
\sigma_3| \leq 1 + \sigma_1 \label{bvcc10}
\end{equation}

\item[(III)] There exist three random variables $A, B, C$ defined
on a single common probability space with the following
properties. The joint probability densities of $(A, B), (A, C)$
and $(B, C)$ are $f_1, f_2$ and $f_3$ respectively. In particular,
the expectation values equal
\begin{equation}
E(A) = E(B) = E(C) = 0 \label{bvcc11}
\end{equation}
the covariances equal
\begin{equation}
E(AB) = \sigma_1 \text{   ,   }E(AC) = \sigma_2 \text{   , }E(BC)
= \sigma_3 \label{bvcc12}
\end{equation}
Using Eqs.(\ref{bvcc10}) and (\ref{bvcc12}) one obtains the six
Bell inequalities for the expectation values $E(AB), E(AC),
E(BC)$.

\end{enumerate}

Proofs: The proof of the Lemma is straightforward. The proof that
conditions (I) and (II) of Theorem1 are equivalent can be done by
inspection. The proof that (III) implies (II) or the six Bell
inequalities obtained from Eq.(\ref{bvcc10}) and Eq.(\ref{bvcc12})
can be carried out by a simple modification of the standard proof
of the Bell inequalities \cite{bell}. Finally, the proof that (I)
implies (III) was outlined at the end of section 3.

We have shown therefore the following. The inequalities of Bell
are a special case of the theorems of Bass and Vorob'ev for the
Ma-EPR experiment. If the 6 Bell inequalities are valid then it is
possible to find three random variables $A, B$ and $C$, defined on
one common probability space that reproduce the three joint pair
densities $f_i$ of Table \ref{TA:ob} and their covariances
$\sigma_1 = E(AB), \sigma_2 = E(AC)$ and $\sigma_3 = E(BC)$. These
covariances satisfy the Bell inequalities. Therefore, if quantum
mechanics predicts that, for a given idealized experiment
involving random variables $A, B$ and $C$ and $E(AB)$, $E(AC)$,
and $E(BC)$, one of the six Bell inequalities in Eq.(\ref{bvcc10})
will be violated or equivalently if the point with coordinates
$(E(AB), E(AC), E(BC)) \in R^3$ does not belong to the covariance
tetrahedron of Fig. \ref{fig:Tetrahedra}, then the random
variables $A, B$ and $C$, that are supposed to form the basis for
the model of this idealized experiment, can not be defined on one
common probability space. We note that the work of Fine
\cite{fine} has already shown the importance of a joint density
and therefore of a common probability space in the derivations of
Bell-type inequalities. The importance of a common probability
space was also stressed more recently in \cite{entrop1} and other
publications.

In summary, we have shown that the definability of $A, B$ and $C$
on one common probability space (OCPS) is a necessary and
sufficient condition for the validity of Bell's inequalities and
that this condition is of a purely mathematical nature and has
nothing to do with the questions of non-locality or counterfactual
reasoning that usually surround discussions of the Bell
inequalities. The condition is, however, related to some of the
physics of EPR experiments in a variety of ways that will be
discussed in section 6.

We add here that other inequalities of similar type such as the
Clauser-Horne-Holt-Shimony (CHHS) \cite{chhs} inequalities can be
treated similarly, although with greater algebraic exertion (16
linear equations in 16 unknowns). Their validity is again a
necessary and sufficient reason that all involved random variables
are defined on one common probability space. In fact, a theorem
analogous to Theorem1 above holds, with the covariance tetrahedron
replaced by a four-dimensional polytope. This polytope equals the
intersection of the four-dimensional parallelepiped, defined by
the four CHHS inequalities, and the four-dimensional cube with
vertices $(\pm 1, \pm 1,\pm 1,\pm1)$. The details will be
published elsewhere.

\section{Bell-type proofs and Bass-Vorob'ev}

In view of the OCPS condition and the theorems of Bass and
Vorob'ev, the proofs for the Bell inequalities as given by Bell
and others become obvious and at the same time lacking physical
justification.

Consider Bell's original proof \cite{bell}. Here Bell assumes that
all random variables $A, B, C$ are in turn functions of a single
random variable $\Lambda$. Then it is clear that $A, B, C$ are
defined on one common probability space and therefore the
inequalities can not be violated by the pair expectation values as
explained above. It is clear that no $\Lambda$ can exist that
leads to a violation of the inequalities for purely mathematical
reasons as already found by Bass much earlier. Bell's physical
justification is wanting because he attempts to show that the
inequalities follow from the fact that $\Lambda$ does not depend
on the settings ${\bf a}, {\bf b},...$. In fact, it does not
matter on what $\Lambda$ depends as long as the resulting $A, B$
and $C$ are random variables defined on one probability space. We
will discuss this in more detail below.

Other well known proofs \cite{leggett} invoke ``counterfactual"
reasoning of the following kind: If, for example, $A$ is measured
given a certain information that we denote by $\lambda$ (a value
that $\Lambda$ assumes in a given experiment) and that is carried
by the correlated spin pair, then one could have measured with
another setting, say $\bf b$ and the same $\lambda$. As we have
explained in more detail previously \cite{hpnp}, it is permissible
to ask the question of what would have been obtained if the
measurement had been performed with a different setting. It is
also permissible to hypothesize the existence of an element of
reality related to that different setting if that different
setting had been chosen. However, to assume then, as is always
done in Bell type proofs, that all these possible different
measurement results are actually contained in the data set of
actual outcomes of the idealized experiment is arbitrary and
against all the rules of modelling and simulation especially for
the particular case of the Aspect-type experiment and all other
known EPR experiments \cite{hpnp}. Naturally, we do not have to
pay for all items on a restaurant's menu just because we could
have chosen them. We call this latter assumption the extended
counterfactual assumption (ECA). ECA is equivalent to the
assumption that $A, B$ and $C$ only depend on one random variable
$\Lambda$ and is therefore an assumption, not a proof. As a
consequence, ECA implies that $A, B$ and $C$ are defined on one
common probability space. In view of the Bass-Vorob'ev theorem it
leads to a contradiction from the outset irrespective and
independent of any physical considerations.

\section{EPR-physics and probability spaces}

A number of physical conditions have been given in the past that
have been thought to be necessary and sufficient for the Bell
inequalities to be valid. Most prominently among these conditions
ranks the definition of an objective local parameter space
\cite{peres}, \cite{leggett}. This definition involves several
conditions that are automatically fulfilled in our Kolmogorovian
model as has been outlined before \cite{hpnp}; it further implies
the existence of elements of reality that contain information
related to the spin (represented by the random variable $\Lambda$)
and, most importantly Einstein locality. Armed with the knowledge
that the validity of the Bell inequalities as described above is
equivalent to the assumption that $A, B, C$ can be defined on one
common probability space, we must now ask the question how this
fact can be related to the condition of an objective local
parameter space i.e. essentially to Einstein locality and the
existence of elements of reality.

We first deal with the question of the relation between the
elements of reality that are ``carried" by the correlated spin
pair and the elements $\omega$ of a probability space. Part of the
work around the Bell theorem concentrates on the question whether
elements of reality that determine (or at least co-determine) the
outcome of the spin measurement can exist. Is not $\omega$ such an
element of reality and do we not assume then its existence to
start with? The answer is that the $\omega$'s represent only a
necessary tool to count and average all measurements correctly.
Whether or not the outcome of a single measurement is the causal
consequence of an element of reality is, at this point, not
discussed. The symbol $\omega$ represents just the choice of the
goddess Tyche (Fortuna) for the given experiment. Of course, if an
element of reality exists, $\omega$ can just represent this
element. The question of whether such elements of reality can
exist in nature and do explain the EPR experiments was, of course,
a subject of the Einstein-Bohr debate and is also subject of our
discussion here. To explore this question using the Bell
inequalities we need to explore whether there exist physical
reasons that demand the definition of $A, B, C$ on one probability
space.

\subsection{Physical reasons for definition on one probability space
for source parameters only}

A very important and broadly applicable physical reason for the
definition of $A, B, C$ on one common probability space arises for
the case in which all random variables are characterized only by
the information emanating from a common source. If in addition
this information is stochastically independent of the settings
(delayed choice arguments), then in line with our notational
convention $A, B, C$ are completely determined by one random
variable $\Lambda$ corresponding to the elements of reality
$\lambda$. These elements of reality can be viewed as the value
the random variable $\Lambda$ assumes for the experiment that we
denoted by $\omega$ i.e. we have the relation $\Lambda(\omega) =
\lambda$. The settings may, of course, also be treated as random
variables and may be defined on a separate probability space.
However, because $\Lambda$ and the settings are stochastically
independent, all random variables can be defined on one common
probability space namely the product space. We have discussed
details of these facts in \cite{hpnp}. Under these conditions the
Bell inequalities will hold and the mathematical model obeying
these conditions is in contradiction to quantum mechanics. We will
show in the next section how this contradiction can be resolved by
still using a classical space-time framework and just adding time
and setting dependent equipment random variables in addition to
the source random variable $\Lambda$. We would like to emphasize,
however, that even though the system consisting of source
parameters only correctly can be ruled out, this fact does not
necessarily have anything to do with Einstein locality. For
example, we can introduce a source parameter represented by a
random variable $\Lambda_1$ that operates only if the settings
${\bf a}, {\bf b}$ and $\bf c$ are employed and $\Lambda_1$
``knows" of these settings by action at a distance. Similarly we
admit a completely different source parameter $\Lambda_2$ that
operates and operates only if the three different settings ${\bf
d}, {\bf e}$ and $\bf f$ are going to be chosen. Again $\Lambda_2$
``knows" of these settings ${\bf d}, {\bf e}$, $\bf f$ by action
at a distance. As long as $\Lambda_1$ is a random variable defined
on some probability space and $\Lambda_2$ is a random variable
defined on some possibly different probability space, the Bell
inequalities formed as before for the settings ${\bf a}, {\bf b}$,
$\bf c$ respectively for ${\bf d}, {\bf e}$,$\bf f$ are valid in
spite of the assumption of action at a distance.

Thus a contradiction exists between the results of quantum
mechanics and the physical assumptions that have just been
described and that appear, on the surface, to be very general.
This contradiction has therefore been explained by some authors
invoking violations of Einstein locality \cite{bellbook}. Others
have given more reasonable, albeit noncommittal, explanations by
postulating that (i) the elements of reality simply do not exist
and/or (ii) there exists a ``contextuality" as discussed in
\cite{peres}. Different contexts of measurements provide then
different probability spaces. There were also other choices to
explain the difficult situation such as (iii) counterfactual
reasoning was held responsible for the difficulties \cite{peres}.
As we have shown, no counterfactual reasoning is necessary to
derive the inequalities and the extended counterfactual reasoning
(ECA) described above is flawed from the viewpoint of mathematical
modelling. We will show in the next section that explanations (i)
and (ii) can, in principle, be reformulated in such a way as to
have a natural explanation in the space-time of relativity. We
note that (i) and (ii) contain in essence Bohr's interpretation:
the spin is determined in the moment of measurement and, with
respect to measurements in any of the two wings of the experiment,
there is essentially the question of ``an influence on the very
conditions which define the possible types of prediction..."
\cite{bohr}.

\subsection{A space-time interpretation of Ma-EPR that agrees
with Bohr in essence}

\begin{figure}[htbp]
    \centering
        \includegraphics[width=0.30\textwidth]{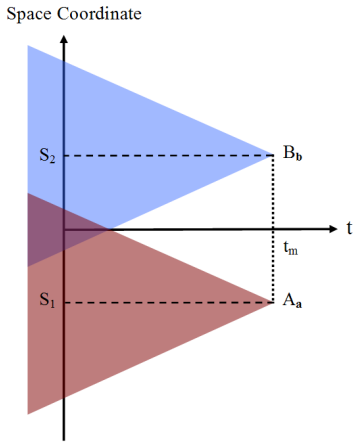}
    \caption{Light cone figure}
    \label{fig:LightCones}
\end{figure}

As the basis for our reasoning in this section, we will assume or
postulate certain properties for the parameters and random
variables of the probability theory that are in harmony with
special relativity. We define with each basic experiment that
corresponds to an element $\omega$ of the probability space a pair
of light-cones corresponding to locations and time at which the
experiments are performed as shown in Fig \ref{fig:LightCones}.

The elements of reality and the corresponding random variables of
the mathematical model are permitted to be functions of the
space-time coordinates of the respective light-cones. As parameter
random variables we admit not only source parameters but also
equipment parameters for each measurement station. What we
introduce below is a dependence of the equipment parameters of a
given station on the setting vector in the light-cone of that
station and an additional dependence on the time of measurement of
a clock in the inertial frame of the equipment.

All we need to achieve is to derive a model for Ma-EPR within the
space-time of relativity that is not refuted by Bell-type
inequalities and agrees with (i) and (ii) in spirit (if not the
letter). For this it is only necessary to find an Einstein local
model with random variables $A, B, C$ that can not be defined on
one common probability space. To show that this is possible we
revert to the standard notation using the settings as subscripts
and denoting the functions in the two experimental wings by
$A_{\bf a}, A_{\bf b}$ on one side and $B_{\bf b}, B_{\bf c}$ on
the other. We continue to permit all functions to be functions of
a source parameter $\Lambda$ which may have a time dependence e.g.
$\Lambda$ may depend on the time of emission of the correlated
pair. However, we also add equipment random variables. Of course,
equipment parameters have been discussed before in many research
articles. But none of them considered the role of time
dependencies of these equipment parameters except our work (see
\cite{hpnp}). We permit that the probability densities of these
additional random variables depend on the time of measurement
$t_m$ as shown by a local clock and also to depend on the local
setting. We indicate this latter fact by denoting the additional
random variable e.g. for setting $\bf a$ by $\Lambda_{\bf a}(t_m)$
we then have $A_{\bf a} = A_{\bf a}(\Lambda, \Lambda_{\bf
a}(t_m))$ and similar for the other settings and the $B$'s on the
other side e.g. $B_{\bf b} = B_{\bf b}(\Lambda, \Lambda_{\bf
b}(t_m))$. Notice that all light-cones for different measurement
times may contain different $\Lambda_{\bf a}(t_m)$ even though the
setting is the same. No matter how a probabilistic model is
conceived, different light-cones can certainly support different
probability distributions for the elements of reality. Assume now
that as in the Aspect-type experiment the settings on each side
are sequentially changed. Because according to relativity this
change of settings to take place requires a time interval of
length bounded away from 0 by a positive constant $c_0$, all
light-cone pairs of a sequence of measurements are different and
each such experiment may be on a different probability space with
a different density of the involved random variables. Furthermore,
let $\eta$ be an element of a probability space that determines
the random times of measurement i.e. $t_m(\eta)$ is the actual
measurement time of a given experiment. We now show that physical
reasons, derived from the framework of relativity only,
necessitate the involvement of different probability spaces if one
postulates the existence of time and setting dependent Einstein
local equipment parameters.

Theorem2: Assume that there exist a source parameter $\Lambda$ and
equipment parameters $\Lambda_{\bf a}, \Lambda_{\bf b}$ and
$\Lambda_{\bf c}$ such that $\Lambda_{\bf a}, \Lambda_{\bf b}$ and
$\Lambda_{\bf c}$ not only depend on the setting vectors ${\bf a},
\bf b$ and $\bf c$, respectively, but also on the time $t_m$ of a
given measurement. Here we consider $t_m$ to be a random variable
$t_m = t_m(\eta)$. Thus
\begin{equation}
\Lambda_{\bf a} = \Lambda_{\bf a}(t_m(\eta)) \text{   ,
}\Lambda_{\bf b} = \Lambda_{\bf b}(t_m(\eta)) \text{   ,
}\Lambda_{\bf c} = \Lambda_{\bf c}(t_m(\eta)) \label{ccc1}
\end{equation}
The source parameter $\Lambda = \Lambda(\omega)$ is permitted to
depend on emission time. We assume further that the random
variables corresponding to the measurements of spin $A_{\bf a},
A_{\bf b}, B_{\bf b}$ and $B_{\bf c}$ all equal to $\pm 1$ are
functions of the source parameter and of the equipment parameters
$\Lambda_{\bf a}, \Lambda_{\bf b}$ and $\Lambda_{\bf c}$. Then,
under the assumption that the velocity of light in vacuo is an
upper limit for the velocities by which the settings can be
changed, there is no probability space on which all of
\begin{equation}
A_{\bf a} = A_{\bf a}(\Lambda(\omega), \Lambda_{\bf a}(t_m(\eta))
\nonumber
\end{equation}
\begin{equation}
A_{\bf b} = A_{\bf b}(\Lambda(\omega), \Lambda_{\bf b}(t_m(\eta))
\nonumber
\end{equation}
\begin{equation}
B_{\bf b} = B_{\bf b}(\Lambda(\omega), \Lambda_{\bf b}(t_m(\eta))
\nonumber
\end{equation}
\begin{equation}
B_{\bf c} = B_{\bf c}(\Lambda(\omega), \Lambda_{\bf c}(t_m(\eta))
\label{ccc2}
\end{equation}
can be consistently defined.

Proof: Let $I$ be any time interval of length $|I| \leq \frac {1}
{2} c_0$. Let $M$ be a measurable set in the range of $\Lambda$
and let $F, G$ and $H$ be sets in the ranges of $\Lambda_{\bf a}$,
$\Lambda_{\bf b}$ and $\Lambda_{\bf c}$ respectively. Then
\begin{equation}
[(\omega, \eta): t_m(\eta) \in I, \Lambda(\omega) \in M,
\Lambda_{\bf a}(t_m(\eta)) \in F, \Lambda_{\bf b}(t_m(\eta)) \in
G, \Lambda_{\bf c}(t_m(\eta)) \in H] \label{bvcc13}
\end{equation}
is the impossible event and therefore has probability 0. Recall
that $\omega$ signifies the sending out of a particular particle
pair from the source. This result simply reflects the
impossibility in the space-time of relativity to accomplish two
different settings on both sides within the same short time
interval and all for the same $\omega$. Hence for each time
interval $I$ each of the sixteen probabilities
\begin{equation}
P[(\omega, \eta): t_m(\eta) \in I, A_{\bf a}(\cdot) = \pm 1,
A_{\bf b}(\cdot) = \pm 1, B_{\bf b}(\cdot) = \pm 1, B_{\bf
c}(\cdot) = \pm 1] = 0 \label{bvcc15}
\end{equation}
must vanish. Here $(\cdot)$ denotes the dependence on source and
equipment parameters that in turn depend on $\omega$ and $\eta$
respectively just as in Eq.(\ref{ccc2}). Now let $J$ be a finite
but arbitrarily long time interval. Then $J$ can be split up into
a large but finite number $N$ of intervals $I_i, i = 1,2,...,N$
with length $|I_i| \leq \frac {1} {2} c_0$. Then the probability
in Eq.(\ref{bvcc15}) with $I$ replaced by $J$ also must vanish
because of finite additivity and thus $A_{\bf a}, A_{\bf b},
B_{\bf b}$ and $B_{\bf c}$ cannot be defined on a common
probability space as claimed.

In other words, not only must we have different probability spaces
involved in the Aspect-type experiment for mathematical reasons,
we must have different probability spaces for physical reasons,
the requirements of relativity. We emphasize that none of the
assumptions in the above proof imply any synchronization of the
measurement times with certain settings. Both settings and
measurement times can be chosen randomly, only the measurement
times in $S_1$ and $S_2$ are correlated for any given photon pair.

Note that the essence of Bohr's discussion is not violated by the
above. We just need to view both spin and measurement equipment in
the sense of information theory: the measurement outcome is really
not the single consequence of the source information $\lambda$
that characterizes particle properties but also that of the
measurement equipment and the corresponding $\lambda_{\bf a}(t_m)$
etc.. These equipment parameters correspond to the use of decoding
machines in information theory \cite{shannon}. Both the source
information content together with that of the decoding machines or
equipment parameters (that involve different probability spaces)
determine the measurement outcomes i.e. the values $\pm 1$ that
the functions $A_{\bf a}$ etc. assume. In a larger sense this
fulfills the spirit of Bohr. The spin does not really exist before
the measurement, but only in the very moment of measurement is the
outcome determined (decoded) and can not be separated from the
equipment and act of measurement. The contextuality is implicitly
contained in the dependence of the probability densities of the
various variables on measurement time. For example, it is now
incorrect to say that it makes no difference if one measures with
setting $\bf b$ or setting $\bf c$ on the other side. It does make
a difference because one necessarily makes these different
measurements during different time intervals. The measurements in
both wings are also performed at the same clock-time or at least
at correlated clock-times which opens the possibility of
correlations between the two wings even though the settings are
randomly chosen.

What is the meaning then of the Aspect et al. \cite{eprex}
experiment in view of the above discussions? If one assumes that
this experiment is free from any problems related to non-ideal
experimental conditions and if one assumes that a space-time
explanation must be possible then the Aspect et al. experiment has
proven the existence of setting and time dependent equipment
parameters.

\section{Conclusion}

We have shown that the inequalities of Bell can be derived as
special cases of a more general theorem found by Bass ten years
earlier. We have further shown that the Bell inequalities are
valid if and only if the three random variables involved can
actually be defined on a common probability space. As a
consequence the Bell theorem is correct at least for the following
systems of hidden variables, in the sense that these systems can
be ruled out:

\begin{enumerate}

\item{} Source parameter $\Lambda$ only,

\item{} Source parameter $\Lambda$ and equipment parameters $\Lambda_{\bf a}$,
$\Lambda_{\bf b}$ and $\Lambda_{\bf c}$ that depend only on the
respective settings.

\end{enumerate}

On the other hand, equipment parameters that depend on the
measurement times as well as on the respective instrument settings
can not be ruled out. A space-time explanation of the Aspect et
al. experiment is therefore not ruled out by Bell's inequalities.
Any such space-time explanation can not rely on source parameters
only but must involve a certain type of time and setting dependent
equipment parameter random variables. Thus, the validity of the
Bell inequalities for objective local parameter spaces has not
been proven by any of the proofs reported in the literature
\cite{bellbook}, \cite{leggett}, \cite{peres}.

\section{Acknowledgement}

The authors would like to thank M. Aschwanden for creating the
figures of the manuscript and helpful suggestions. Support of the
Office of Naval Research (N00014-98-1-0604) is gratefully
acknowledged.

\end{document}